\documentclass{jpsj-suppl}

\title{An Antiproton Deceleration Device for the GBAR Experiment at CERN}

\author{A. \textsc{Husson}$^{1}$ and D. \textsc{Lunney}$^{1}$}

\inst{$^{1}$CSNSM, Universit\'e de Paris-Sud, IN2P3/CNRS, Orsay, France}

\email{audric.husson@csnsm.in2p3.fr; david.lunney@csnsm.in2p3.fr}

\recdate{June 23, 2016}

\abst{The GBAR experiment aims at performing the first free-fall experiment with antihydrogen atoms in order to test the weak equivalence principle with antimatter. Antihydrogen ions are synthesized through a double charge exchange reaction and laser cooled to energies small enough to see gravitational effects. Anti-ion synthesis requires a large number of antiprotons, hence high efficiency. We report here the design of an new electrostatic deceleration device providing low energy antiprotons of a few keV. This technique will avoid losses inherent to the use of a degrader foil.}

\kword{antiproton, antihydrogen, deceleration, electrostatic optics}

\begin{document}
\maketitle

{\em Contribution inadvertently omitted from the Proceedings of Low Energy Antiproton Physics (LEAP), held  6-11 March 2016 in Kanazawa, Japan} \\

\section{Introduction}
Numerous theoretical and experimental investigations prove that the gravitational mass and the inertial mass of ordinary matter are equivalent to one part per $10^{12}$\cite{WEP1,WEP2}. 
Pointing out the lack of evidence considering the gravitational behaviour of antimatter systems, theorists have built new models that in some cases violate CPT or Lorentz invariance as an explanation for baryogenesis or cosmological problems.
But physicists are used to say that absence of proof is not a proof of absence. 

With advances in antimatter synthesis, tests of the Weak Equivalence Principle (WEP) are now achievable.  
Considering the weakness of gravity compared to any residual electrical or magnetic force, direct free-fall experiments with charged antimatter appear unrealistic.
Antihydrogen is the best candidate because of its electrical neutrality and its long lifetime compared to other neutral anti-systems like $Ps$. 
But this apparent advantage implies the use of complicated system to trap neutral atoms. 
Such trapping techniques have already been performed by antimatter experiments like $ATRAP$\cite{ATRAP1, ATRAP2, ATRAP3} and $ALPHA$\cite{ALPHA1, ALPHA2, ALPHA3}.

Even if these collaborations have already managed to synthesize antihydrogen atoms, the measurements of $F=M_g/M$, defined as the ratio of the gravitational mass $M_g$ to the inertial mass $M$ of antihydrogen, remains quite difficult.
According to Ref.\cite{ALPHAgrav}, this factor $F$ has been bounded by the ALPHA collaboration between -65 and 110, far from the $|F|=1$ values.

Accepted by the CERN Research Board in 2012\cite{GBARspsc}, GBAR (Gravitational Behaviour of Antimatter
at Rest) aims at performing the first free-fall experiment using $\bar{H}$ atoms from an at-rest state and to estimate the $\bar{g}$ acceleration on Earth. 
To reach energies low enough to see gravitational effects, GBAR will first synthesize antihydrogen ions using laser-excited positronium. 
These ions are then captured and cooled to few neV. 
The production of such ions relies on a double charge-exchange reaction (see Eq \eqref{eq1} - \eqref{eq2}) for which antiproton energies between 1 and 6 keV are predicted to be most favourable \cite{Comini}. 
Recent calculations of antihydrogen production using positronium \cite{Charlton} show that orders of
magnitude can be potentially gained by cooling antiprotons below room temperature.

\begin{samepage}
\begin{align}
	\label{eq1}
	\bar{p} + Ps &\rightarrow \bar{H} + e^-
	\\
	\label{eq2}
	\bar{H} + Ps &\rightarrow \bar{H}^+ + e^-
\end{align}
\end{samepage}

The new extra low-energy antiproton ring ELENA \cite{ELENA}, currently under construction, will extend
the deceleration capacity of the CERN Antiproton Decelerator (AD) providing 100 keV antiproton
beams for experiments devoted to antihydrogen, including GBAR. 
If 100 keV is still quite high for reaction (\ref{eq1}), this regime allows the use of electrostatic devices. 
We report here the design of a new electrostatic device to further decelerate antiprotons from 100 keV to about 1 keV which avoids losses from the use of energy-degrading foils and the complexity of
intermediate trapping.

\begin{figure}[!b]
\includegraphics[trim = 60mm 110mm 95mm 40mm, width=\textwidth]{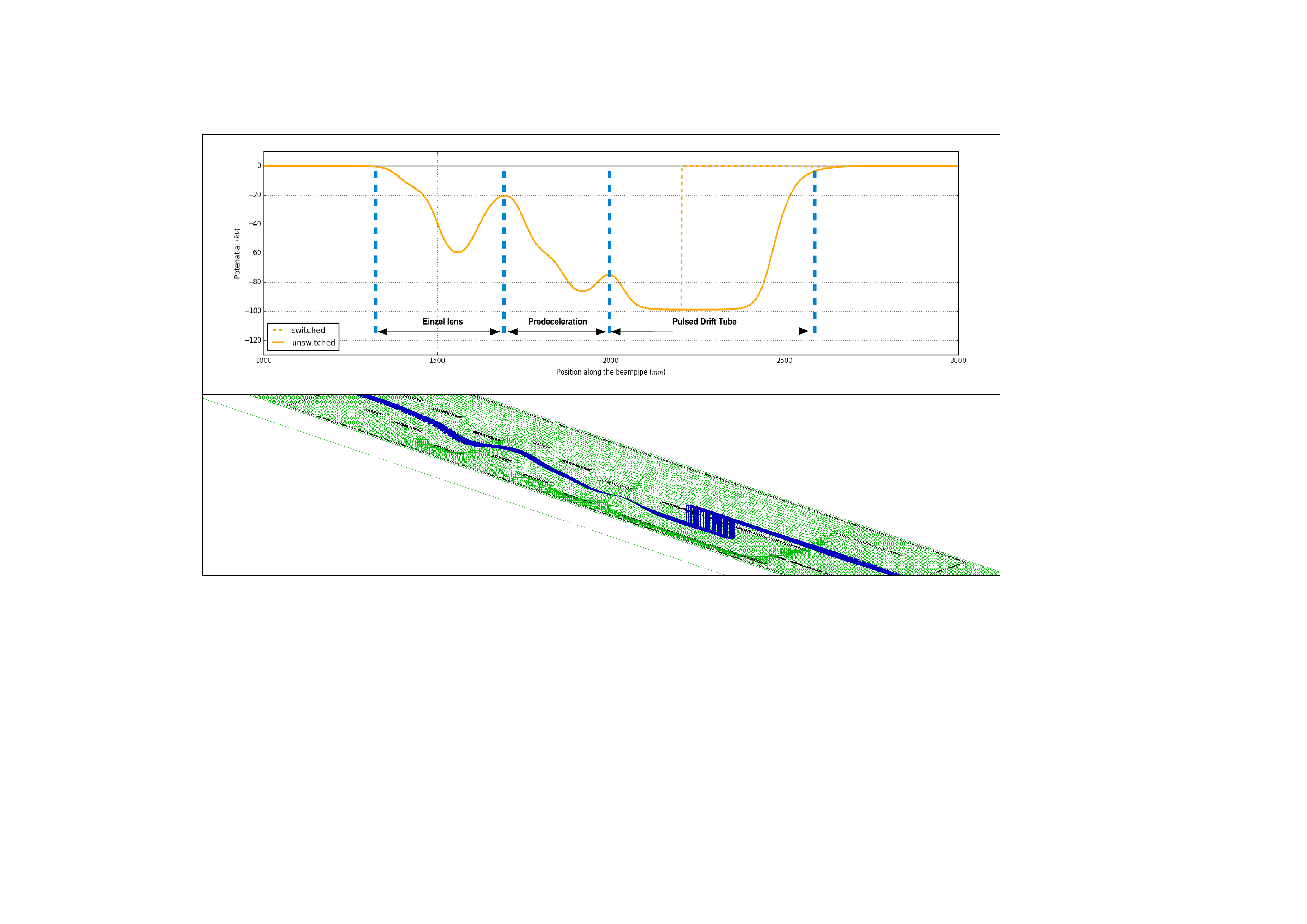}
\caption{ \textit{Top} :Potential profile along the decelerator system axis. The different regimes of the deceleration are delimited by the blue dashed lines. \textit{Bottom} : SIMION view of the deceleration of a 100 keV antiproton bunch to 1 keV. The green mesh corresponds to the potential map. The blue lines represent the trajectories of the particles with respect to the potential. The step in the trajectories is when the PDT is switched.}
\label{potential}
\end{figure}

\section{Principle}

We propose a decelerating system made of a series of six electrostatic, cylindrical lenses followed by a $40$ cm-long drift tube held at -99 kV.
The number, the size and the gap between each electrode have been chosen in order to minimize the risk of sparks.
The cylindrical symmetry (axisymmetry) guarantees good resistance to high voltage by limiting coronal discharges for a relatively low cost. 

Transported at ground potential, the antiproton pulse slows down facing the strong electric field at the entrance of the drift tube. 
Passing though it, particles are not influenced by the applied potential which is switched from -99 kV to ground while they are inside. 
When exiting the drift tube, antiprotons do not see a re-accelerating potential and retain their residual 1 keV energy\cite{Lunney}. 

The electric field gradient at the entrance of the pulsed drift tube (PDT) leads to a strong focusing of the beam. 
The 5 deceleration electrodes upstream are set to potentials ensuring a parallel transport through the tube. 
These potentials have been simulated with the SIMION software. 
The best configuration found has been divided in three stages (see Figure-\ref{potential}) using a strong focusing at $30$ mm upstream of the drift tube. 
The three first electrodes play then the role of an Einzel lens where the central electrode is held at $-60$ kV. 
In a second time, a step-by-step increase in the voltage applied on the electrodes from $-10$ kV to $-90$ kV.
This ramp decelerates the bunch and avoids the explosion directly at the entrance of the main cylinder.

Bunches coming from the new ELENA ring are expected to be $300$ ns long. 
While decelerating, the antiproton pulse will be spatially compressed from roughly $1.3$ m for a $100$ keV beam to $20$ cm inside the pulsed drift tube. 
Only conservative forces are involved except space-charge effects therefore a small increase of a few percent in the bunch length is foreseen.

The main limitation for such an electrostatic system appears to be the momentum spread $\Delta p/p$. 
Indeed, variations in the energy momentum induce changes in the divergence of the most energetic particles susceptible to be kicked out the beam path by edge effects close to the electrodes. 
According to our simulations, a worst case $\Delta p/p$ value of $10^{-2}$ would correspond to a loss of $33\%$ of the antiprotons. 
In the framework of the ELENA/GBAR collaboration with a momentum spread $\Delta p/p\ =10^{-3}$ at $100$ keV, a transmission rate through the decelerator of nearly $95\%$ is expected.

\section{Experimental setup}

A prototype, working with a 50 keV beam, has been tested at CSNSM (Orsay, France) to pursue the study on the deceleration principle. 
This experimental apparatus is composed of a Penning discharge source whose acceleration potential can be set to 50 kV. 
The resulting energy of the exiting particles is $48$ keV with an energy spread of $250$ eV. 
This spread is mainly due to the discharge potential applied to generate ions and to the fluctuating gas pressure inside the plasma chamber. 
The geometrical emittance of such a source is roughly $25\ \pi$ mm.mrad at 48 keV. 
In addition to $H^+$, $H^+_2$, $N^+$ and $H_2O^+$ ions are produced (see Figure-\ref{tof}-\textit{left}). 
Only $H_2O^+$ ions are used as deceleration probes. 
The other species are kept unchanged as a time-of-flight reference at the end of the line. 

\begin{figure}[!b]
	\includegraphics[trim = 0mm 37mm 24mm 0mm,width=\textwidth]{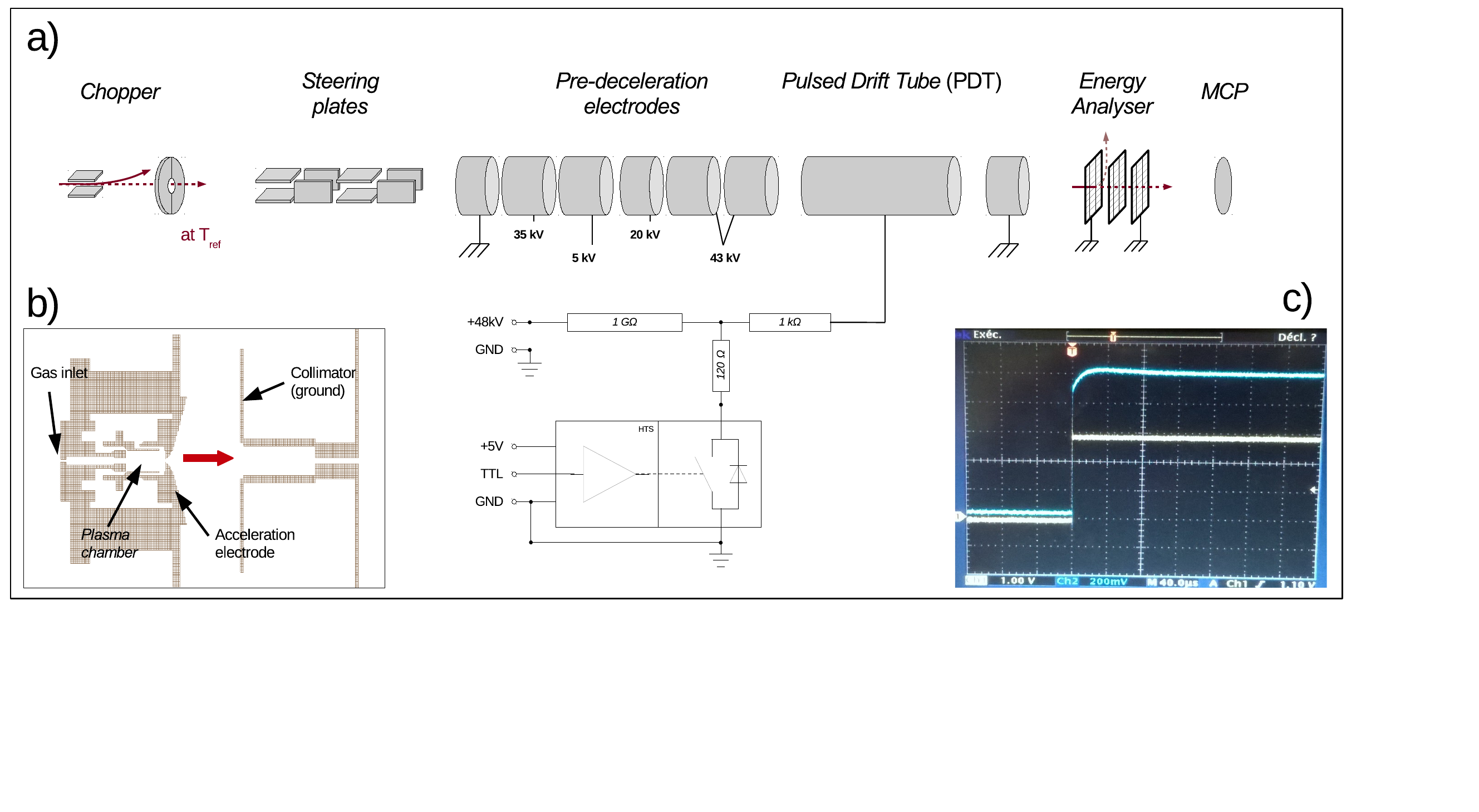}
	\caption{\textit{a)} Schematic of the deceleration apparatus - \textit{b)} Simulation file of the Penning source (SIMION) - \textit{c)} Oscilloscope screen showing the potential on the PDT when switching at 10kV (\textit{the yellow line is the trigger signal, the blue line is the potential on the drift tube}).}
	\label{apparatus}
\end{figure}

The beam is chopped with a pair of deflector plates followed by a $15$ mm aperture. 
The applied voltage is disrupted by a Push-Pull GHTS 10kVDC-100A BEHLKE switch with a $7$ second period.
The initial time $T_{ref}$ is defined when the potential on the deflector plates is turned off. 
A minimum relevant bunch length of $2\ \mu s$ has been used to reduce any distortion phenomena induced by a too short impulsion on the electronic trigger of the transistor switch.
Finally, the beam is centred through the deceleration electrodes with 4 pairs of steering plates, two for each direction.

The drift tube is pulsed by a MOSFET fast high-voltage transistor switch (Push MOSFET HTS Behlke $140\ kVDC-200A$) synchronized on the chopper cycle with a delay time of $1.8\ \mu s$. 
Specific high-voltage ceramic resistors allow to adjust the rising and the falling time of such a capacitive circuit. 
For a $47$ kV potential, falling-times of order $150$ ns were recorded while rising-times were broadened on half of a period to limit local charge effects when the tube is re-polarized. 

For technical reasons, we weren't able to use a refocusing system after the pulsed drift tube. 
We then chose to install the diagnostic part directly downstream of the PDT with enough distance to differentiate the different species in time and to not suffer the electromagnetic noise while switching the tube. 
Although the refocusing problem was partly solved by increasing the potential on the upstream electrodes, the beam signal suffered a too low intensity downstream the PDT for diagnostic. 

The beam energy was scanned with the use of a energy analyzer made of three successive meshes. 
The outermost grids are grounded while the intermediate one is held at a floating voltage. 
Without refocusing lens, the transmission rate through this apparatus is evaluated to $36\%$ taking into account the Gaussian beam profile and the geometry of the scanner. 

Finally, a MCP detects the beam and sends the signal through a low-pass filter and a amplification chain to an oscilloscope coupled with a monitoring computer. 

\section{Results}

Even without refocusing, the transmission is still quite acceptable: 
$48\%$ of the beam remains downstream the drift tube and the energy scanner. 
This value is better than expected, probably because the beam profile is slightly different than the gaussian profile we have used.
By stopping the chopper cycle, we confirmed that the observed signal was not electronic noise due to the pulse on the drift tube. 

\begin{figure}[!b]
	\includegraphics[trim= 0cm 22cm 0.50cm 0cm, width=\textwidth]{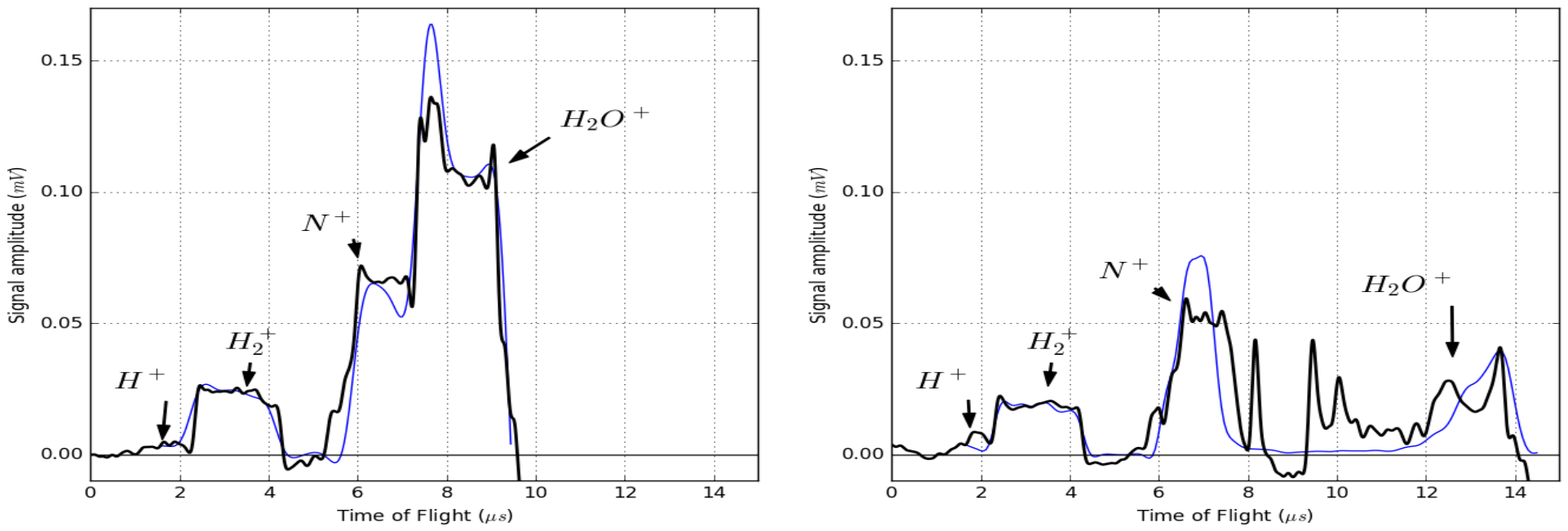}
	\caption{Time of flight spectra from the discharge source. The blue lines are the predicted signals from SIMION simulations. (\textit{Left}) - $48$ keV beam without switching. (\textit{Right}) - After the deceleration process. The $H^+$, $H_2^+$ and $N^+$ ions remain at $48$ keV while the $H_2O^+$ ions are partly decelerated to $1$ keV slowing their time of flight.}
	\label{tof}
\end{figure}

During the data acquisition, a first TOF spectrum was achieved with the PDT polarized at $47$ kV and another one by switching the PDT from $47\ kV$ to ground. 
Between both sets of measurements, the potential on the pre-decelerating electrodes was adjusted to compensate the optic changes and maximize the signal on the MCP. 

As Figure-\ref{tof} shows, species are clearly discriminated on the time-of-flight spectra before and after the beginning of the deceleration system. 
Considering the distance $d=3.4$ m between the chopper and the MCP, the obtained spectra are in good agreement with the predicted values based on SIMION simulations.

Almost $35\%$ of the remaining $H_2O^+$ ions are decelerated, but those between $10$ and $12\ \mu s$ are not completely decelerated to $1\ keV$. 
This can be explained by the difficulty to synchronize the PDT switching on the passage of the $H_2O^+$ ions in the tube. 
A shorter bunch or a longer time-of-flight could improve this measurement.

\begin{figure}[!]
	\centering
	\includegraphics[trim = 0mm 220mm 5mm 0mm,width=\textwidth]{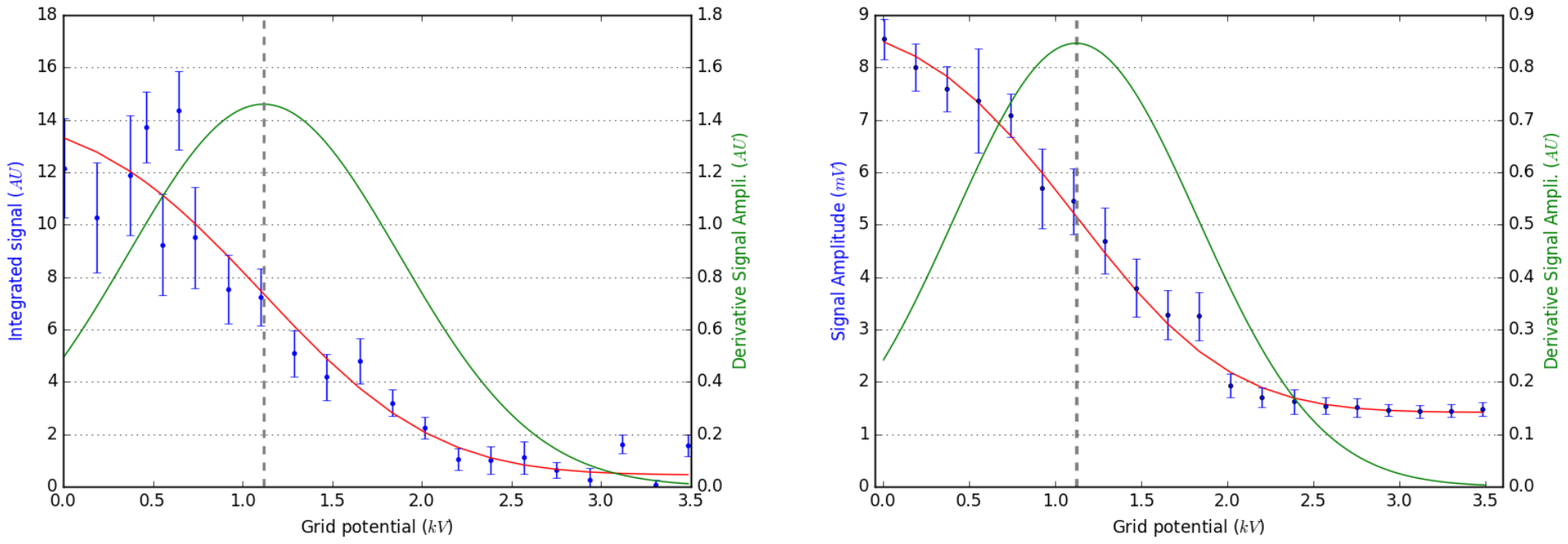}
	\caption{Energy spectra after the deceleration process. The blue points are the data affected by their statistical error for 100 samples. Red lines are error function fit curves. Green lines correspond to the derivative of the fit curves. Derivatives follow a gaussian distribution\  $D = A.\exp(-(x-\mu)^2/(2 \sigma^2))$.}
	\label{results}
\end{figure}

During a second test, the remaining energy of the $H_2O^+$ ions was investigated. 
The beam was directed through the energy analyser. 
The potential on it was raised step by step until the disappearance of the last peak on the TOF spectrum. 
The potential configuration of the source, the PDT and on the pre-deceleration electrodes were kept unchanged.

Both the amplitude and the integrated signal of the $H_2O^+$ peak were registered and fitted with an error function. 
The derivative of the fit curve gives us the energy profile of the decelerated ions. 
The centroid values, $\mu _{ampl} = 1'119$ V and $\mu _{integ} = 1'116$ V found, confirm an effective deceleration from $48$ keV to  $1.1\pm0.7$ kev. 

\section{Conclusion}

We designed and tested a new decelerating device for antiproton bunches in the framework of the GBAR collaboration. 
This system is supposed to complete the deceleration of pbar beams coming from the new ELENA ring at CERN.
Preliminary tests with positive ions at low energy have demonstrated that deceleration is feasible with good efficiency. 
The electrical behaviour of the different pulsed high voltage components was also successfully tested. 
A final version working at $-100$ kV will be installed during the 2016-2017 CERN Year-End Technical Stop (YETS).

\end{document}